\documentclass[a4paper,11pt]{article}
\usepackage{pos}

\newcommand{\radshock} {\texttt{radshock}}
\newcommand{\komrad} {\texttt{Komrad}}

\title{Radiation-mediated shocks in GRB prompt emission}

\author*[a, b]{Filip Alamaa}
\author[b]{Felix Ryde}
\author[b]{Christoffer Lundman}

\affiliation[a]{Institut d'Astrophysique de Paris, Sorbonne Universit\'e and CNRS, \\
UMR 7095, 98 bis bd Arago, F-75014 Paris, France}

\affiliation[b]{Department of Physics, KTH Royal Institute of Technology, and The Oskar Klein Centre, \\
SE-10691 Stockholm, Sweden}

\emailAdd{filip.alamaa@iap.fr}

\abstract{The debate regarding the emission mechanism in gamma-ray bursts has been long-standing. Here, we study the spectral signatures of photospheric emission, accounting for subphotospheric dissipation by a radiation-mediated shock. The shocks are modeled using the Kompaneets RMS approximation (KRA). We find that the resulting observed spectra are soft, broad, and exhibit an additional break at lower energies. When fitting a collection of 150 mock data samples generated by the model, we obtain a distribution of the low-energy index $\alpha$ that is similar to the observed one. These results are promising and show that dissipative photospheric models can account for many of the observed properties of prompt gamma-ray burst emission.}

\FullConference{High Energy Phenomena in Relativistic Outflows VIII (HEPROVIII)\\
 23-26 October, 2023\\
Paris, France\\}

\begin{document}
\maketitle

\section{Introduction}
The prompt emission mechanism in gamma-ray bursts (GRBs) remains elusive despite many decades of debate. One promising avenue is photospheric emission, which is due to radiation escaping once the initially optically thick fireball transitions to optically thin. Due to the ultra-relativistic velocity of the jet, this emission appears to the observer as a short duration pulse of $\sim 1~$MeV photons \citep{Paczynski1986, Goodman1986}. Early models of photospheric emission predicted that the observed spectrum should be similar to that of a blackbody. However, it has become clear that quasi-blackbody spectra cannot explain the observations, since observed GRB spectra are generally much broader \citep[e.g.,][]{Poolakkil2021}. 

One way to broaden the observed spectrum is via energy dissipation below the photosphere \citep{ReesMeszaros2005}. Here, we consider subphotospheric energy dissipation via radiation-mediated shocks (RMSs), which has been suggested by many authors \citep[e.g.,][]{Eichler1994, LevinsonBromberg2008, Bromberg2011b, Levinson2012}. We focus on the spectral signatures of such emission and compare them to observations.

\section{The Kompaneets RMS Approximation}
\begin{figure*}
\begin{centering}
    \includegraphics[width=0.328\columnwidth]{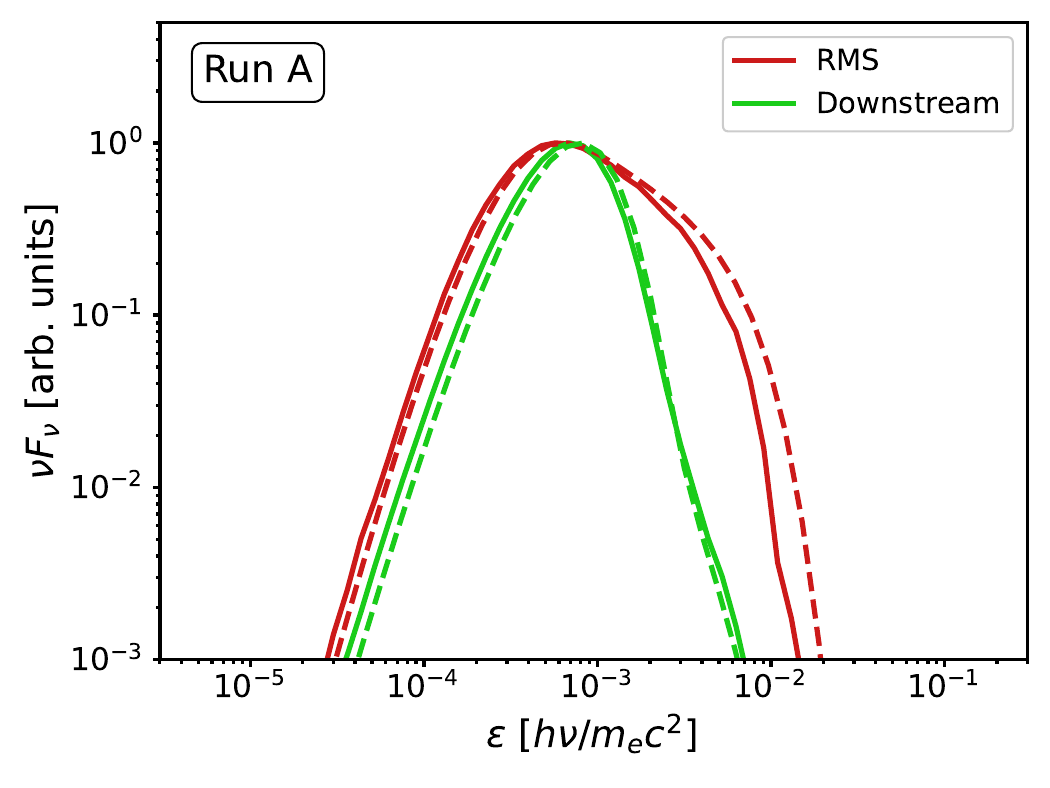}
    \includegraphics[width=0.328\columnwidth]{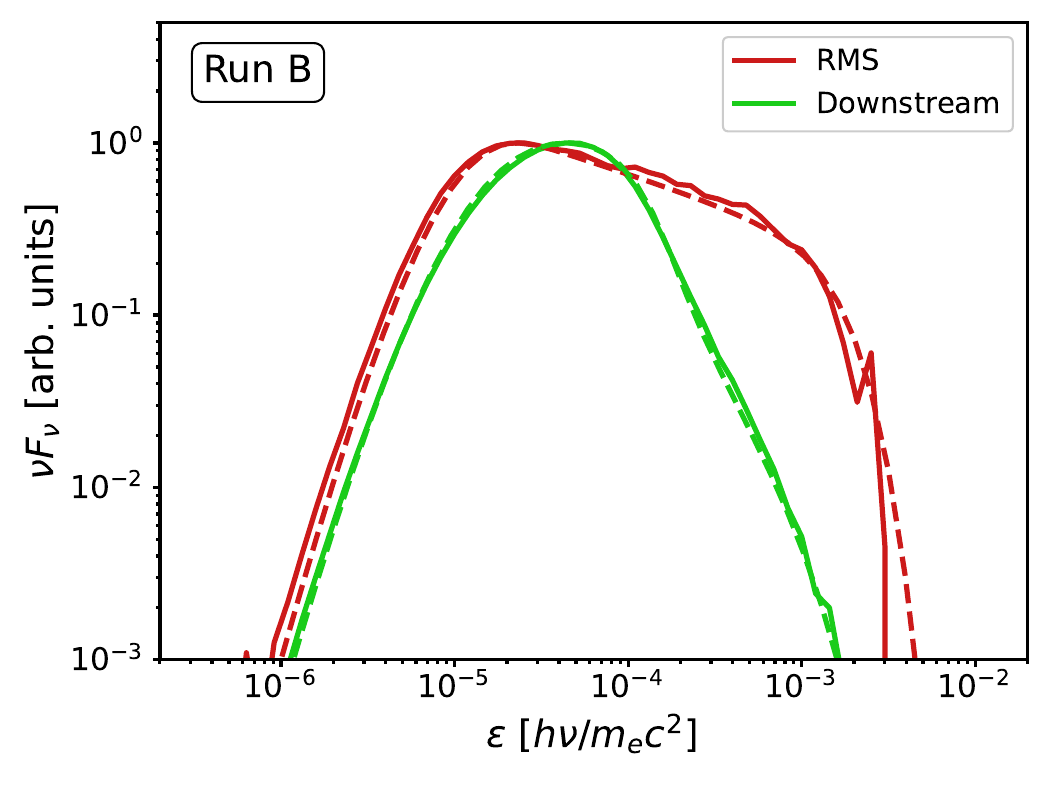}
    \includegraphics[width=0.328\columnwidth]{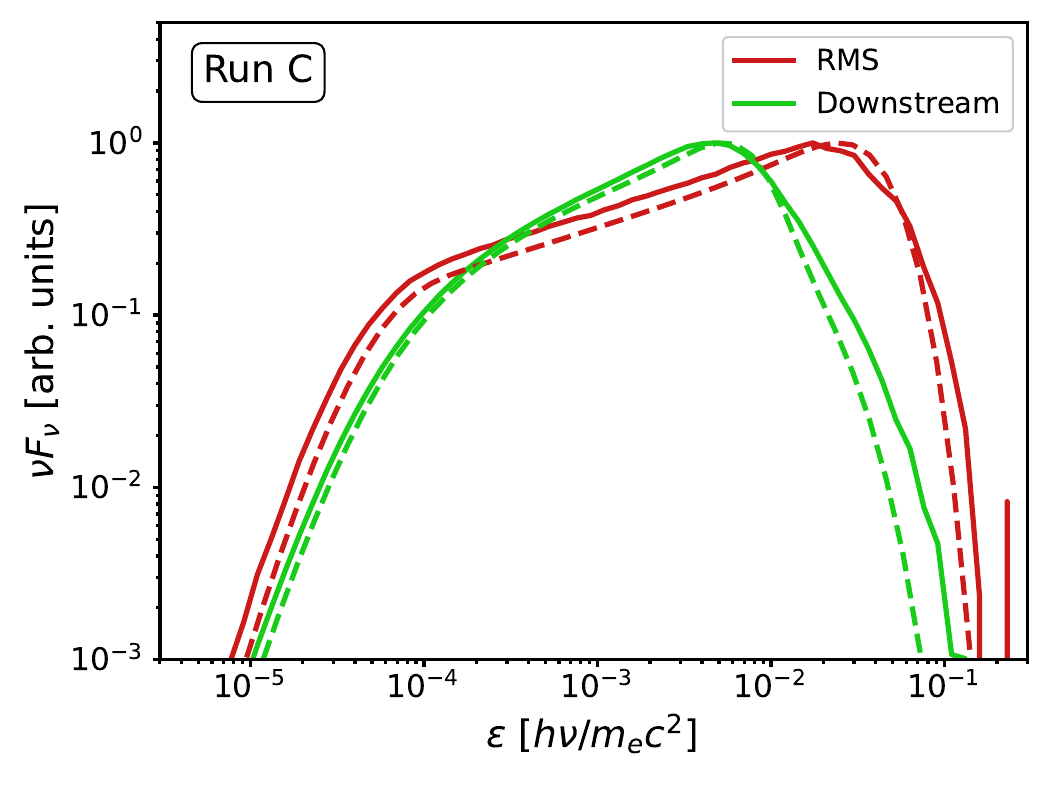}
    \caption{Comparison between the full scale RMS simulation \radshock\ (solid lines) and the approximation \komrad\ (dashed lines) for three different runs. Red lines show the photon distributions in the shock region, while the green lines show the photon distributions downstream of the shock. Figure originally appeared in \citep{Samuelsson2022}.}
    \label{Fig:RunAC}
\end{centering}
\end{figure*}

We model the RMS using the Kompaneets RMS Approximation \citep[KRA,][]{Samuelsson2022, SamuelssonRyde2023}. The numerical code that implements the KRA is called \komrad. The KRA is valid for photon rich RMSs in regions of negligible magnetic fields and for mildly relativistic or slower RMSs. These conditions are likely met in the subphotospheric region of GRBs \citep{Piran1999, Bromberg2011b, LevinsonNakar2020}. Within this domain, the KRA has been compared to the full-scale radiation hydrodynamic simulation code \radshock \ \citep{Lundman2018} and found to give very good agreement (see Figure \ref{Fig:RunAC}). \komrad\ takes $\sim 4$ orders of magnitude less computational time to run compared to the full-scale radiation hydrodynamic simulation. This massive time reduction allows us to study RMSs quantitatively (see Section \ref{Sec:distributions}). Furthermore, \komrad\ is quick enough for the model to be well suited to fitting GRB data and the first ever RMS model fit for GRBs was performed on a time resolved spectrum in GRB 150314A in \citep{Samuelsson2022}.
 
\section{Evolution of a typical spectrum}\label{Sec:spectral_evolution}
\begin{figure*}
\begin{centering}
    \includegraphics[width=0.4\columnwidth]{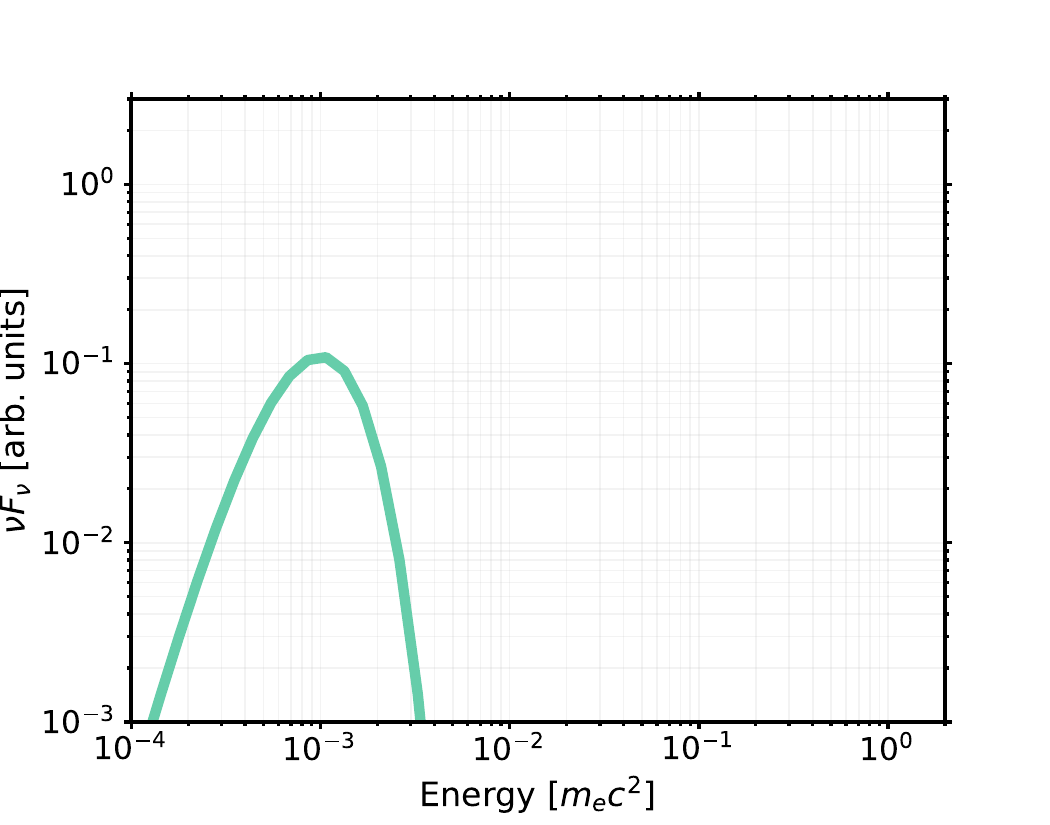}
    \hspace{5mm}
    \includegraphics[width=0.4\columnwidth]{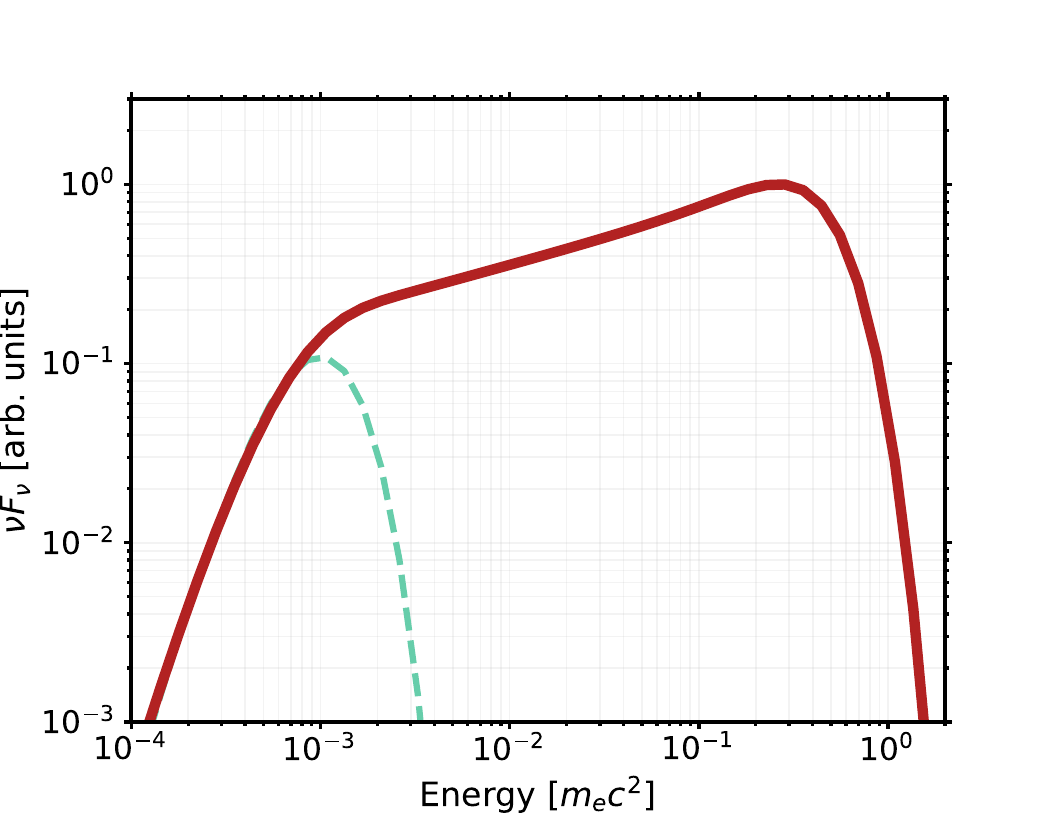}\\
    \includegraphics[width=0.3\columnwidth]{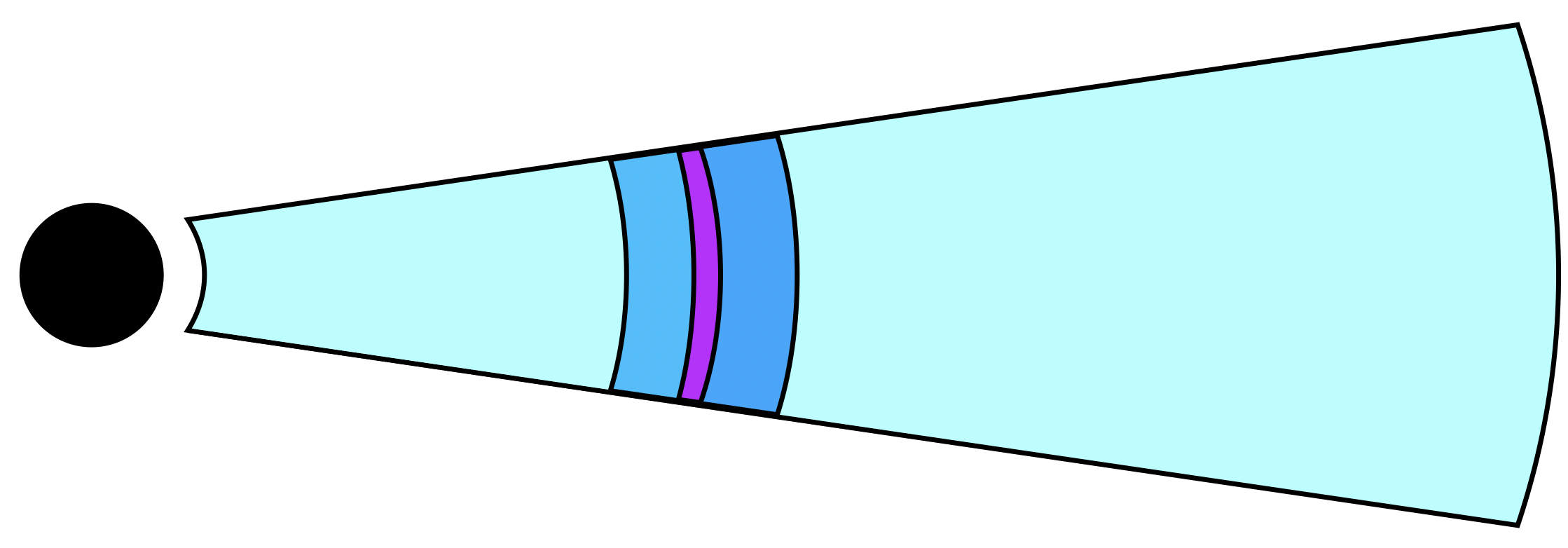}
    \hspace{20mm}
    \includegraphics[width=0.3\columnwidth]{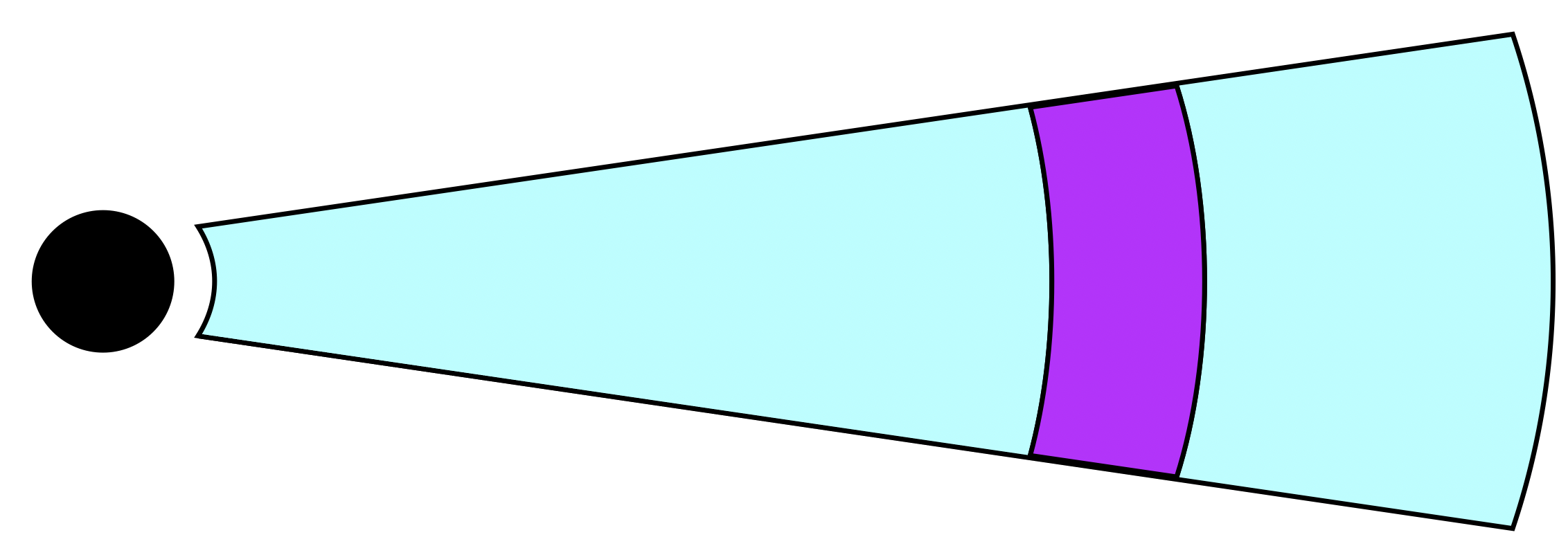}\\
    \includegraphics[width=0.4\columnwidth]{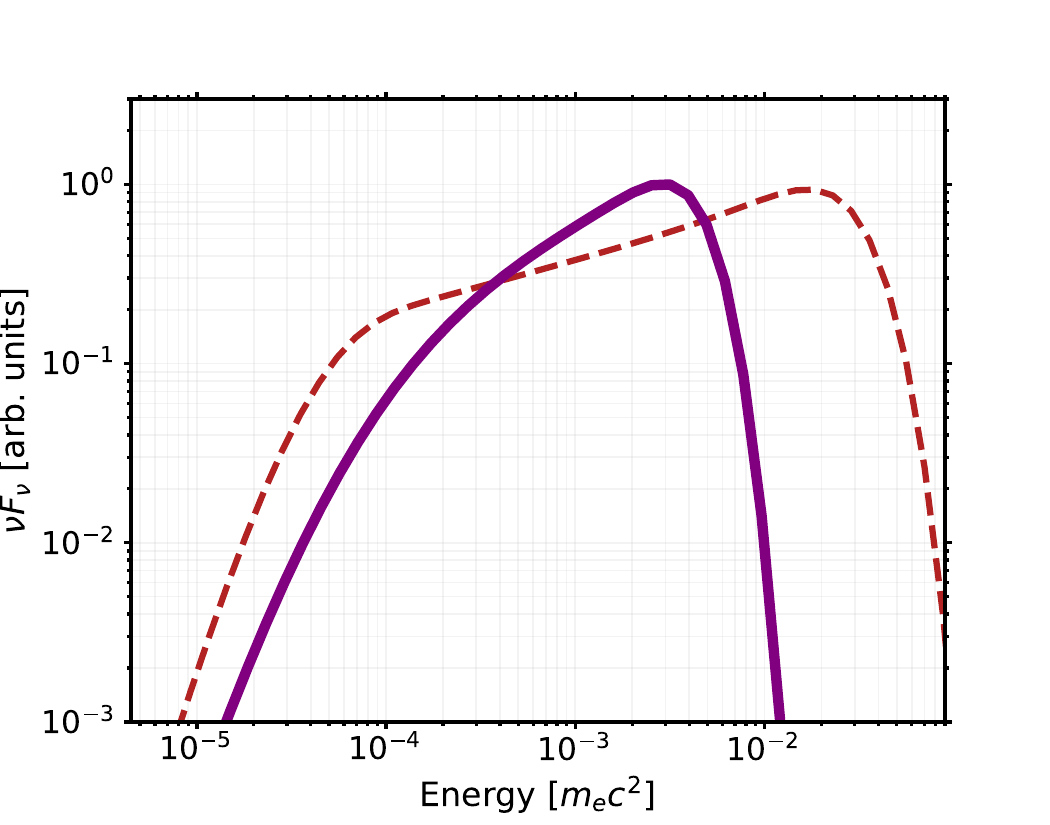}
    \hspace{5mm}
    \includegraphics[width=0.4\columnwidth]{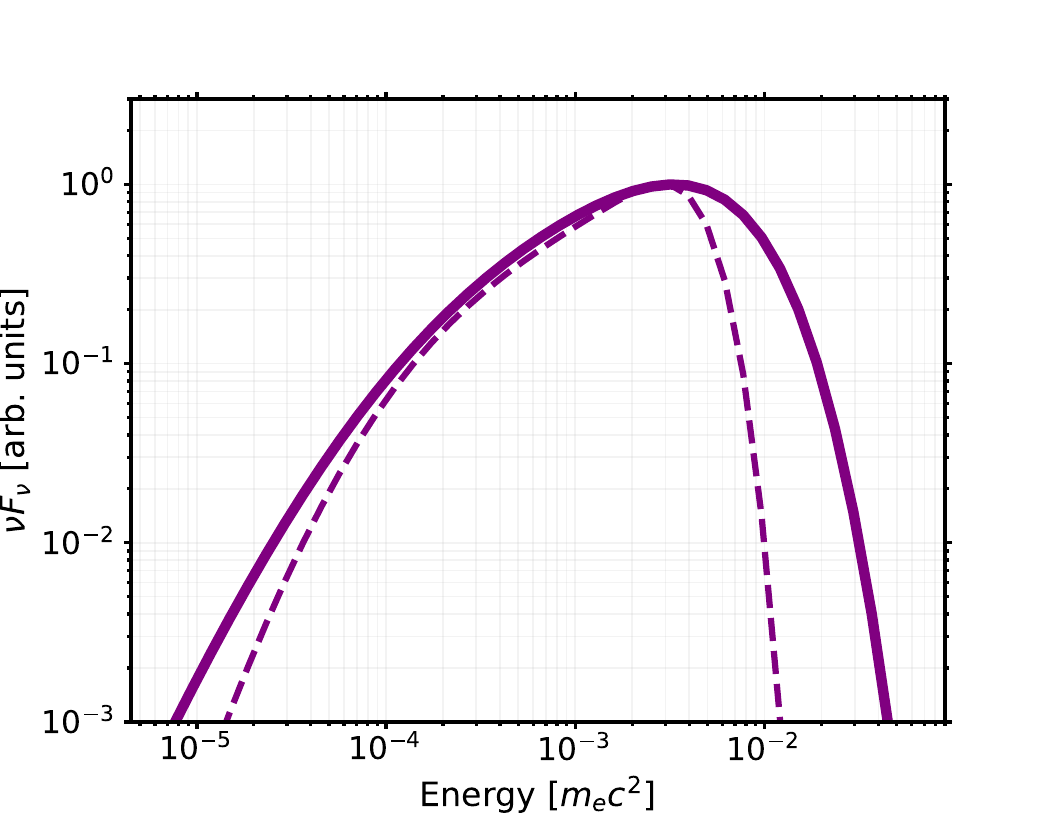}\\
    \hspace{5mm}
    \includegraphics[width=0.37\columnwidth]{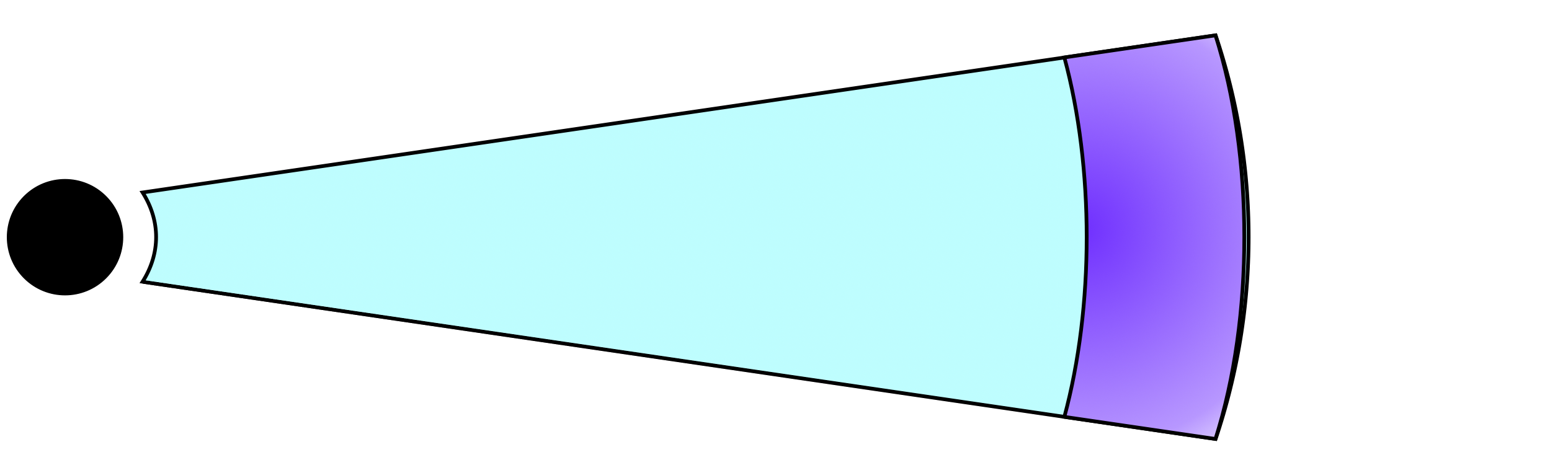}
    \hspace{5mm}
    \includegraphics[width=0.37\columnwidth]{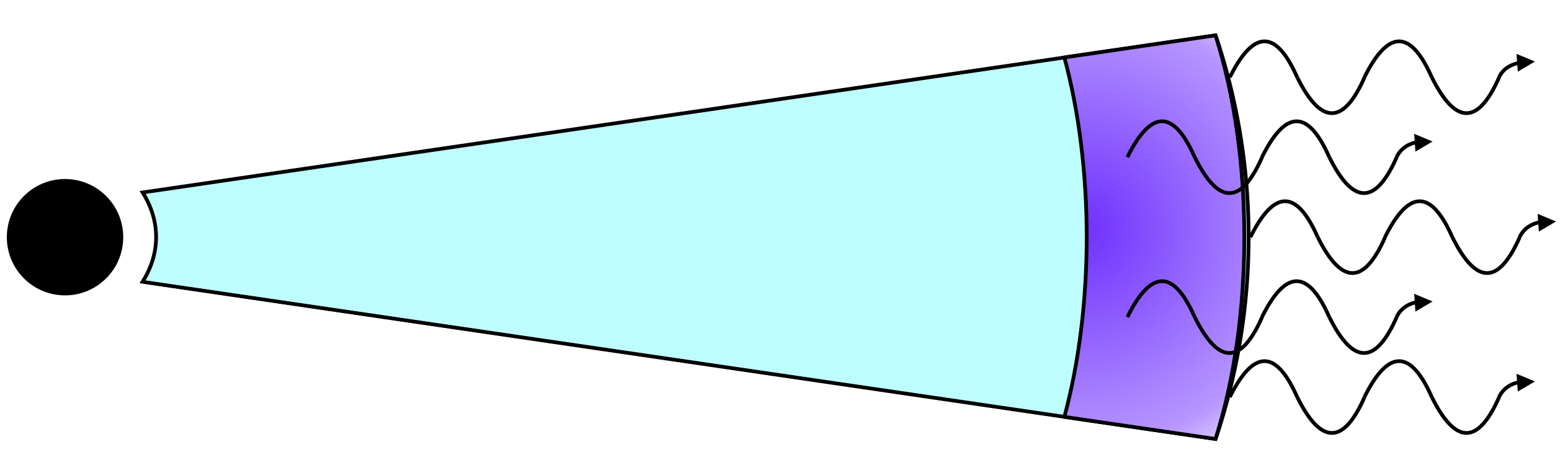}
    \caption{Four different stages in the evolution of the photon distribution from initial thermal upstream (top left) to the emitted spectrum (bottom right). Below each panel is a schematic of the GRB jet at each time. Details about the evolution and each panel can be found in Section \ref{Sec:spectral_evolution}.}
    \label{Fig:spectral_evolution}
\end{centering}
\end{figure*}

To get a feeling for the evolution of the photon distribution from thermal upstream to observed spectrum, we show four different stages of the evolution in four different panels in Figure \ref{Fig:spectral_evolution}. Here, the RMS is assumed to result from a subphotospheric internal collision between two shells. Below each panel is a schematic showing the jet at the corresponding time, with the photosphere being at the rightmost edge of the cartoon jet.

In the first stage (upper left panel) the quicker of the two shells has just caught up with the slower one. The spectrum shows the initial unperturbed upstream distribution, which in the simulation is taken to be a thermal Wien distribution. Two shocks start propagating into the two blobs, all the while the whole ejecta is moving outwards ultra-relativistically. If the masses and comoving densities are similar for the two shells, the two resulting shocks have similar properties \citep{Samuelsson2022}. This allows us to model only one of the two shocks.

Photons are continuously injected from the upstream into the shock region, where they are energies via bulk Comptonization. Additionally, photons are advected from the shock region into the downstream. This creates a power-law spectrum within the RMS \citep{Lundman2018, Ito2018}. The spectrum within the RMS at the moment where the RMS has crossed the entire upstream is shown by a solid red line in the upper right panel. The green dashed line shows the initial distribution for comparison. 

Since the shock has crossed the upstream before the jet has become optically thin, the photon distribution must still be tracked up until the photosphere. The photons in the downstream are allowed to interact with the thermal population of electrons, leading to the high-energy photons losing energy while the low-energy photons gain energy. Thus, the photon distribution relaxes towards a thermal equilibrium. However, if the optical depth of the collision is not too high ($\tau \lesssim \textrm{few} \ 100$), the photon distribution will not have time to fully thermalize before the ejecta reaches the photosphere. The partially thermalized downstream spectrum at the photosphere is shown in the bottom left panel by a solid purple line. Furthermore, the whole spectrum has decreased in energy as a consequence of adiabatic cooling. The shock spectrum from the upper right panel, accounting for adiabatic cooling, is shown by a dashed red line for comparison. 

Finally, we account for the fact that the observed spectrum consists of a superposition of comoving spectra, emitted at different optical depths and angles to the line of sight. This leads to a broadening and smoothening of the spectrum. The broadened spectrum is shown by the solid purple line in the bottom right panel, compared to the comoving spectrum shown by the dashed line.

\begin{figure}
\begin{centering}
    \includegraphics[width=0.59\columnwidth]{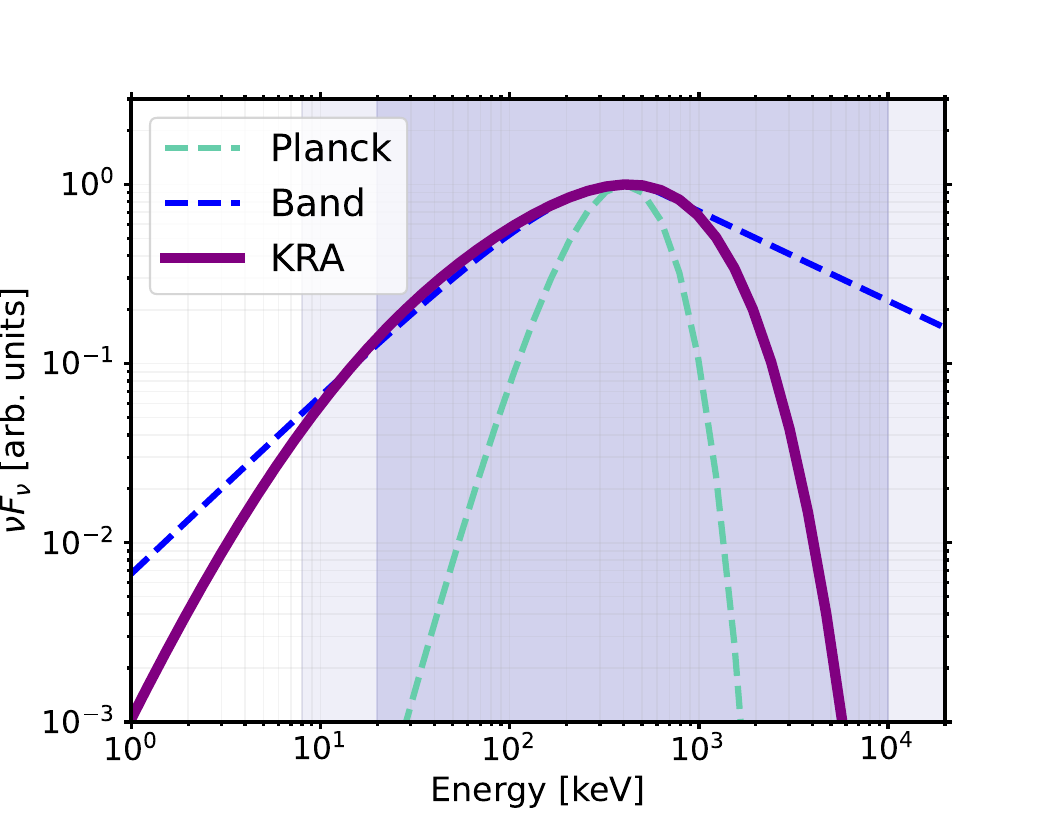}
    \caption{Final observed spectrum, shown compared to a Planck function and a typical Band function as indicated in the legend. The Band function has a low-energy index $\alpha = -1$ and a high-energy index $\beta = -2.5$. The purple shading shows the GBM energy sensitivity, with darker shading indicating higher sensitivity. The observed spectrum is much broader and softer compared to the Planck function. In contrast to the Band function, the KRA spectrum exhibits a hardening at lower energies.}
    \label{Fig:observed_spectrum}
\end{centering}
\end{figure}

In Figure \ref{Fig:observed_spectrum}, the observed spectrum is shown. It is identical to the spectrum shown by the solid purple line in the bottom right panel of Figure \ref{Fig:spectral_evolution}, but Doppler boosted into the observer frame using a bulk Lorentz factor of $\Gamma = 300$ and accounting for redshift with $z=1$. The spectrum is shown compared to a Planck distribution in dashed green and a typical GRB spectrum in dashed blue. The latter is modeled by a Band function \citep{Band1993} with power-law indices $\alpha = -1$ and $\beta = -2.5$. The purple shading shows the sensitivity of the Gamma-ray Bursts Monitor \citep[GBM,][]{Meegan2009} on board the Fermi Gamma-ray Space Telescope, with darker shading indicating higher sensitivity.

From the figure, it is evident that the observed KRA spectrum is much broader and much softer compared to the Planckian. Therefore, it is not correct to say that photospheric emission in GRBs is necessarily hard. Indeed, the low-energy part of the spectrum is similar to the typical low-energy slope $\alpha = -1$, as evident when comparing with the Band function. Additionally, the spectrum exhibits a double power-law behavior at low energies, which is interesting as this has been detected in several bright GRBs \citep{Ravasio2018, Ravasio2019, Burgess2020, Gompertz2023}.

\section{Distributions of $\alpha$ and $E_{\rm peak}$}\label{Sec:distributions}
\begin{figure*}
\begin{centering}
    \includegraphics[width=0.49\columnwidth]{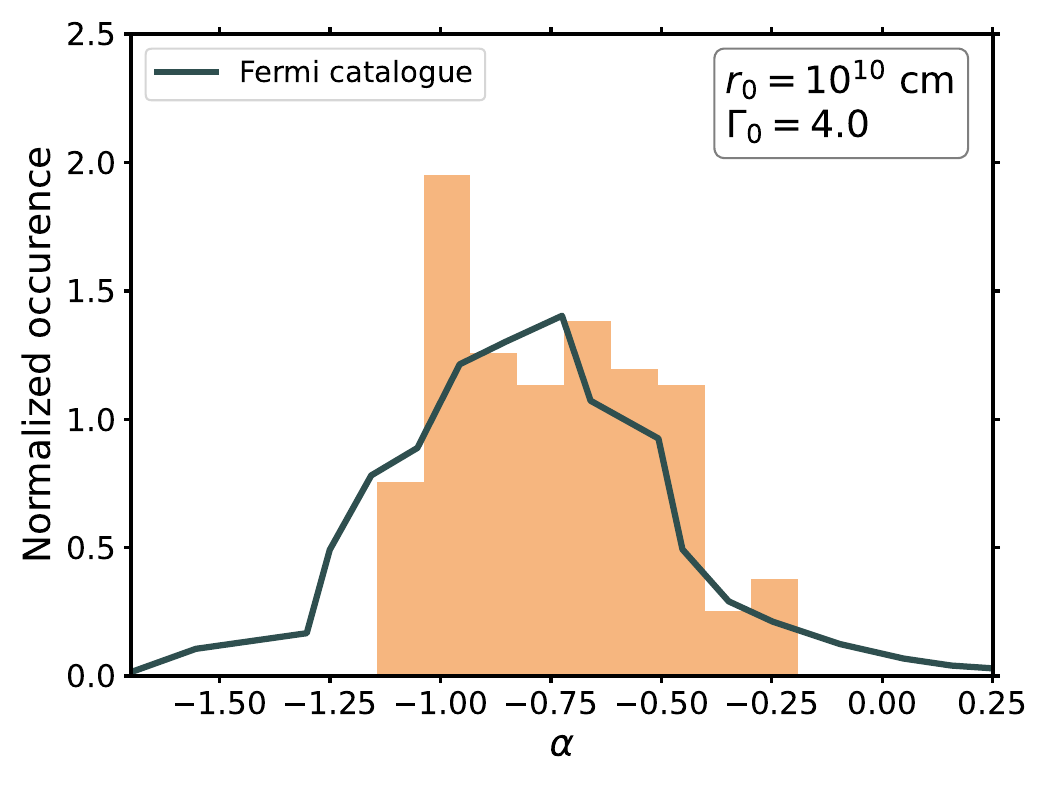}
    \includegraphics[width=0.49\columnwidth]{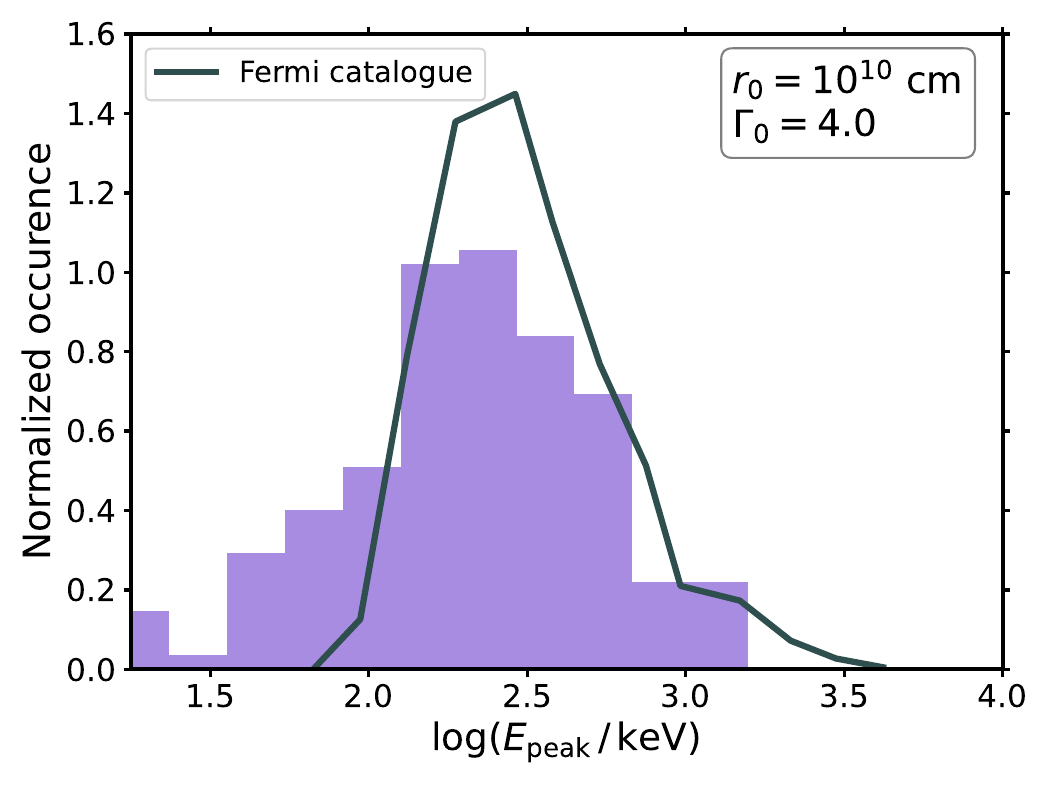}
    \caption{Histograms obtained by fitting 150 mock data sets with a CPL function (see section \ref{Sec:distributions} for details). The left panel shows the distribution of the low-energy index $\alpha$ and the right panel shows the distribution of the peak energy $E_{\rm peak}$. The solid dark green line shows the distributions obtained in the time-resolved Fermi catalogue \citep{Yu2016}. Figure originally appeared in \citep{SamuelssonRyde2023}.}
    \label{Fig:histograms}
\end{centering}
\end{figure*}

To estimate the observational characteristics of the model more quantitatively, we show histograms of the parameter values obtained from 150 fits to mock data sets in Figure \ref{Fig:histograms}. The mock data sets are generated as follows. We consider the internal collision between two blobs with different Lorentz factor $\Gamma_f$ and $\Gamma_s$. The collision radius and strength of the shock is then determined by the ratio $\Gamma_f/\Gamma_s$ \citep{Kobayashi1997, DaigneMochkovitch1998}. We assume the collision occurs at an optical depth $\tau$. These three parameters, together with the initial radius $r_0$ and initial Lorentz factor $\Gamma_0$ at the base of the jet, are sufficient to uniquely determine the observed spectrum \citep[assuming similar mass and comoving density in the two blobs, see][for details]{SamuelssonRyde2023}.  By fixing $r_0 = 10^{10}~$cm and $\Gamma_0 = 4$, and varying $\Gamma_s$, $\Gamma_f$, and $\tau$, we generate 150 mock spectra. These spectra are forward-folded through the GBM response matrix to generate the mock data sets. 

Once the 150 sets of mock data are obtained, we fit each with a cutoff power-law function (CPL). The CPL function is characterized by only two parameters: a power-law index $\alpha$ ($N_E \propto E^{\alpha}$, where $N_E$ is the photon number flux) and a peak energy $E_{\rm peak}$, above which there is an exponential cutoff. The histograms obtained for the two parameters are shown in Figure \ref{Fig:histograms}. The solid dark green line shown the distributions in the latest time-resolved Fermi catalogue \citep{Yu2016}. The obtained distributions are quite similar to the observations, indicating that dissipative photospheric models are promising avenues for the prompt emission in GRBs.

\section{Summary and Conclusion}
We have studied the spectral signatures of photospheric emission including subphotospheric dissipation via an RMS. The RMS is modeled using the KRA, which allow for quantitative studies due to its low computational cost. The observed spectra are broad, soft, and exhibit an additional break at lower energies. So far, the KRA has been used to successfully fit a time resolved spectrum in GRB 150314A \citep{Samuelsson2022} and future work will determine how the model fares against a larger sample of GRBs.

\bibliographystyle{JHEP}
\bibliography{References}

\end{document}